\documentclass[review]{elsarticle}
\usepackage[top=3cm,bottom=3cm,left=3cm,right=3cm]{geometry}
\usepackage{lipsum}
\makeatletter
\usepackage{lineno,hyperref}
\usepackage{textcomp}
\usepackage{amsfonts}
\usepackage{amsmath}
\usepackage{amssymb}
\usepackage{rotfloat}
\biboptions{numbers,sort&compress}
\usepackage{graphicx}
\usepackage{dcolumn}
\usepackage{bm}
\usepackage{array}
\usepackage{setspace}

\def\ps@pprintTitle{%
	\let\@oddhead\@empty
	\let\@evenhead\@empty
	\def\@oddfoot{}%
	\let\@evenfoot\@oddfoot}
\makeatother

\begin{document}
	
	\begin{frontmatter}
		
		\title{Comment on ``Nonlinear charge--voltage relationship in constant phase
			element'' {[}AEU-Int.~J.~Electron.~Commun. 117, 153104 (2020){]}}

		\author{Vikash Pandey}
		\address{School of Interwoven Arts and Sciences, Krea University, Sri City, India}

		\ead{vikash.pandey@krea.edu.in}

\begin{abstract}
In this comment, we show a dimensional inconsistency that plagues
one of the main founding equations, Eq.~$\left(5\right)$, of the manuscript,
{[}Fouda~et~al., AEU-Int.~J.~Electron.~Commun.~117, 153104 (2020){]}.
Also, a resolution of the inconsistency as well as a generalized yet
a better version of the equation are suggested.
\end{abstract}

\end{frontmatter}

In contrast to Newtonian (\textit{ordinary}) calculus, the field of
fractional calculus has long struggled for its justification due to its unusual properties and the lack of a universal interpretation \cite{Podlubny2002}. This is also evident
from the fact that despite having more than three hundred years of
history, the order of the fractional derivative is still mostly obtained
from curve-fitting the experimental data with the theoretically predicted
curves. This comment is presented with the motivation that an utmost care should be given in the application of fractional derivatives in describing
anomalous, complex, and memory-driven physical phenomena \cite{Pandey2016a,Pandey2016,Holm2016}. 

In Ref.~\cite{Fouda2020}, the authors have replaced the multiplication
operation, ``$\cdot$'', between the capacitance and the voltage in the traditional charge--voltage relation of a capacitor, by the convolution operation, \textquotedblleft $\ast$\textquotedblright. This is understandable since the purpose
was to investigate a fractional capacitor that exhibits memory \cite{Westerlund1991,Westerlund1994}.
For the sake of clarity and completeness of the arguments presented here, we
rewrite the Eq.~$\left(5\right)$ from Ref.~\cite{Fouda2020} as:

\begin{equation}
q\left(t\right)=c\left(t\right)\ast V(t),
\end{equation}
where, the authors have regarded, $q$, $c$, $V$, and $t$, as the
accumulated charge, time-varying capacitance, applied voltage, and
time respectively. Using the definition of convolution, if one examines
the units of the left and the right sides of Eq.~$\left(1\right)$, it can be ascertained that the respective
equation is not dimensionally consistent as shown below,
\begin{equation}
q\left(t\right)=c\left(t\right)\ast V\left(t\right)=\int\limits _{0}^{t}\underbrace{c\left(\tau\right)}_{\text{Farad}}\underbrace{V\left(t-\tau\right)}_{\text{Volts}}\underbrace{d\tau}_{\text{Second}}\text{ has units, }C\cdot s.
\end{equation}
As observed from Eq.~$\left(2\right)$ above, the unit on the
right hand side of Eq.~$\left(1\right)$, is, Coulomb $\cdot$ second.
On the contrary, the left hand side of the equation dictates the unit
to be that of a charge, i.e., Coulomb. This is a contradiction.
Any equation, or inequality, corresponding to a physically valid system
must have the same dimensions on both left and right sides, a property
known as dimensional homogeneity. This makes Eq.~$\left(1\right)$
questionable. 

We now suggest the following correction and a comment for the respective
equation. The correction is that, $c\left(t\right)$, is actually
capacitance per unit time and not just a capacitance which the authors
have claimed. Interestingly the authors have indirectly implied the same even
in their paper too. In the paragraph that immediately follows
after Eq.~$\left(5\right)$ in Ref.~\cite{Fouda2020}, the authors mention,
$c\left(t\right)=C\delta\left(t\right)$, where $C$ is the constant
capacitance, and $\delta$ is the Dirac delta function. Here, the argument
for the Dirac delta function is, time, $t$, and since the Dirac delta
function always has the inverse dimension of its argument, this implies
that the units of $c\left(t\right)$ is actually, Farad/second. It must be emphasized that the origin of the dimensionally inconsistent equation, Eq.~$\left(1\right)$, can be traced back to the manuscript that is cited as Ref.~[19] in Ref.~\cite{Fouda2020}, where again the dimensionality of the Dirac delta function has been overlooked.  The suggested correction restores the dimensional homegeneity in Eq.~$\left(1\right)$. 
Incorporating the correction it can be further shown that a better
equation for the charge--voltage relation of a fractional capacitor
is with a differential operator which appears as Eq.~$\left(9\right)$
in Ref.~\cite{Pandey2022}, as follows:

\begin{equation}
q\left(t\right)=C\left(t\right)\ast\frac{dV\left(t\right)}{dt}=\frac{dC\left(t\right)}{dt}\ast V\left(t\right).
\end{equation}
In the special case of a linear memory, i.e., for capacitors whose
capacitance increase linearly with time, both the convolution
relations, Eqs.~$\left(1\right)$ and $\left(3\right)$, yield the
same result, however there are two advantages in using Eq.~$\left(3\right)$ over the corrected
form of Eq.~$\left(1\right)$. First, the one with a differential operator
is applicable for most forms of memory that can be described
as polynomials, be it linear, quadratic, or of an even higher order.
 Second, it has been already established in Ref.~\cite{Pandey2022} that the Eq.~$\left(3\right)$ when interpreted in the light of fractional calculus describes the century-old dielectric
relaxation law, the Curie--von Schweidler law \cite{Curie1889,Schweidler1907,Jonscher1977}. Furthermore, a physical interpretation of the \textit{fractional-order} that appears in the expression for current of a fractional capacitor can be found in  Ref.~\cite{Pandey2022}.


\begin{thebibliography}{10}


\bibitem{Podlubny2002}
I.~Podlubny.
\newblock Geometrical and physical interpretation of fractional integration
  and fractional differentiation.
\newblock {\em Fract.~Calc.~Appl.~Anal.}, 5:367--386, 2002.

\bibitem{Pandey2016a}
V.~Pandey and S.~Holm.
\newblock Linking the fractional derivative and the {L}omnitz creep law to
  non-Newtonian time-varying viscosity.
\newblock {\em Phys.~Rev.~E}, 94:032606, 2016.

\bibitem{Pandey2016}
V.~Pandey and S.~Holm.
\newblock Connecting the viscous grain-shearing mechanism of wave
  propagation in marine sediments to fractional calculus.
\newblock {\em In: 78th
  Eur.~Assoc.~Geoscient.~\&~Engineers (EAGE) Conference and Exhibition}, 2016.

\bibitem{Holm2016}
S.~Holm and V.~Pandey.
\newblock Wave propagation in marine sediments expressed by
  fractional wave and diffusion equations.
\newblock {\em  In: IEEE/OES China Ocean
  Acoustics (COA)}, 1--5, 2016.

\bibitem{Fouda2020}
M.~E.~Fouda, A.~Allagui, A.~S.~Elwakil, S.~Das, C.~Psychalinos, and A.~G.~Radwan.
\newblock Nonlinear charge--voltage relationship in constant phase element.
\newblock {\em AEU-Int.~J.~Electron.~Commun.}, 117:153104, 2020.

\bibitem{Westerlund1991}
S.~Westerlund.
\newblock Dead matter has memory!.
\newblock {\em Phys.~Scripta}, 43:174--179, 1991.

\bibitem{Westerlund1994}
S.~Westerlund and L.~Ekstam.
\newblock Capacitor theory.
\newblock {\em IEEE~T.~Dielect.~El.~In.}, 1:826--839, 1994.

\bibitem{Pandey2022}
V.~Pandey.
\newblock Origin of the {C}urie--von {S}chweidler law and
the fractional capacitor from time-varying capacitance.
\newblock {\em J.~Pow.~Sources}, 532:231309, 2022.

\bibitem{Curie1889}
J.~Curie.
\newblock Recherches sur le pouvoir inducteur sp\'{e}cifique et sur la
  conductibilit\'{e} des corps cristallis\'{e}s.
\newblock {\em Ann.~Chim.~Phys.}, 17:385--434, 1889.

\bibitem{Schweidler1907}
E.~R.~von~Schweidler.
\newblock Studien \"{u}ber die anomalien im verhalten der
  dielektrika (studies on the anomalous behaviour of dielectrics).
\newblock {\em Ann.~Phys.}, 329:711--770, 1907.

\bibitem{Jonscher1977}
A.~K.~Jonscher.
\newblock The ``universal'' dielectric response.
\newblock {\em Nature}, 267:673--679, 1977.


\end{thebibliography}

\end{document}